\documentclass[aps,prl,twocolumn,groupedaddress]{revtex4}
\newcommand{\phiv}{\mbox{\boldmath$\hat{\phi}$}}
\newcommand{\sigmav}{\mbox{\boldmath$\hat{\sigma}$}}
 \newcommand{\rv}{\mbox{\boldmath${r}$}}
\newcommand{\iv}{\mbox{\boldmath$\hat{i}$}}
 \newcommand{\jv}{\mbox{\boldmath$\hat{j}$}}
\newcommand{\kv}{\mbox{\boldmath$\hat{k}$}}
\newcommand{\rhov}{\mbox{\boldmath$\hat{\rho}$}}

\usepackage{graphicx}
\usepackage{epsfig}
\usepackage{dcolumn}
\usepackage{bm}
\usepackage{amsmath}
\usepackage{ifpdf}

\begin{document}


\title{Electromagnetic Spin-Orbit Interactions via Scattering}


\author{L.T. Vuong$^{1}$, A.J.L. Adam$^{2}$, J.M. Brok$^{3}$, M. A. Seo$^{4}$, D. S. Kim$^{4}$, P.C.M. Planken$^2$, H.P. Urbach$^2$}
\email[]{LTV2@cornell.edu}

\affiliation{1 School of Applied and Engineering Physics, Cornell University, Ithaca, NY, USA}
\affiliation{2 Optics Research Group, Delft University of Technology, The Netherlands}
\affiliation{3 Department of Mathematics and Computer Science, Eindhoven University of Technology, The Netherlands}
\affiliation{4 School of Physics and Astronomy, Seoul National University, Korea}


\date{\today}

\begin{abstract}
The longitudinal components of orthogonal-circularly polarized fields carry a phase singularity that changes sign depending on the polarization handedness.  The addition of orbital angular momentum adds to or cancels this singularity and results in polarization-dependent scattering through round and square apertures, which we demonstrate analytically, numerically, and experimentally.  By preparing the incident polarization and arranging the configuration of sub-wavelength apertures, we produce shadow-side scattered fields with arbitrary phase vorticity.  
\end{abstract}
\maketitle
Within the last decade, there have been wide observations of optical phase singularities in the evanescent fields produced by propagating \cite{NFMeasure} and scattered \cite{NFSing} light.  These phase singularities indicate where the electric field is strictly zero and carry information about the electric field Poynting vector and angular momentum \cite{PIO}. A three-dimensional electric field produces three different types of polarization phase singularities \cite{Berry2}, the evolution of which is studied in a rich array of literature \cite{Berry3}.  Our understanding of phase singularities allows us to probe and characterize materials, surfaces, and light propagation dynamics, as well as control and manipulate microparticles \cite{microparticle}.  

Recently, there have been reports that the locations of near-field phase singularities produced by chiral ``gammadion'' \cite{ChiralHole1} and spiral grating structures \cite{ArchSpiral} depend on incident polarization handedness.  These phase singularities are connected to the extraordinary transmission of light through sub-wavelength slits \cite{ChiralHole2}, where whirlpool-like power flows and singularities in the Poynting vector are shown to exist \cite{WhirlPart,LenstraPRL}.  Moreover, azimuthally and radially-polarized vortices, beams with different polarization singularities, are transmitted through apertures with different efficiencies \cite{Leuchs}.  Despite numerous measurements and observations of phase singularities in the near field, the polarization-dependent transmission that occurs at sub-wavelength structures is not fully understood, so that light-metal interactions are neither fully optimized nor controlled. 

Here, we show that polarization-dependent phases at subwavelength-structured materials are concisely described by electromagnetic spin-orbit interactions, where ``spin'' refers to the circular polarization of electric fields, and ``orbital angular momentum'' identifies vortex phasefronts that are the signature of phase singularities.  Electromagnetic spin angular momentum and orbital angular momentum are coupled specifically when the longitudinal electric field component plays a role in the dynamics and previously cited examples of optical spin-orbit interactions investigate oblique reflection or refraction \cite{Zeldovich, Chiao} or high-numerical aperture focusing \cite{HiNAPRL}.  

We explain, for the first time, that polarization-dependent singularities in longitudinal field components dictate which modes and to what extent light is transmitted through thin-film apertures via the process of electromagnetic scattering.  Our results suggest that waveguiding due to finite-thickness materials \cite{Leuchs} is not the only explanation for understanding polarization-dependent transmission through round subwavelength apertures.  Numerical simulations verify our analytical predictions and we reconstruct experimental THz field measurements with subwavelength-resolution that demonstrate polarization-dependent phase singularities.  Our new insight of coherent light-metal interactions enables us to produce shadow-side longitudinal fields with arbitrary phase vorticity by controlling the polarization and configuration of sub-wavelength apertures.  

The topological charge or phase winding associated with the longitudinal or $z$-component of a spin-polarized electric field is $m_z = m_l + m_s$, where $m_l$ is the topological charge associated with orbital angular momentum in the transverse electric field components, and the topological charge associated with the photon spin number, $m_s = \pm 1$, depends on the orthogonal-spin polarization $\sigma_\pm$.   The longitudinal component topological charge is illustrated by representing a continuous-wave circularly-polarized field with transverse mode amplitude $A(\rho, \phi,z) e^{i m_l \phi}$ as
 \begin{eqnarray}
 \tilde{\bf E}= A [e^{i m_l \phi}\sigmav_\pm + \Delta_\pm \kv]e^{i(\omega t - k z)},
  \end{eqnarray}
 where the circular polarization unit vector is represented in cartesian and cylindrical coordinates as ${\sigmav_\pm} =  ({\iv} \pm i{\jv})/{\sqrt{2}}=   ({\rhov}  \pm i   {\phiv})e^{\pm i \phi}/{\sqrt{2}}$.  The longitudinal component of the electric field is calculated by Maxwell's equation $\widehat{\nabla} \cdot \tilde{\bf E} = (\widehat{\nabla}_\perp + \widehat{\partial}_z)\cdot \tilde{\bf E}=0$,
\begin{eqnarray}
\Delta_\pm e^{- i k z}= -\int_{-\infty}^{z} e^{-i k z'} ({\bf \nabla}_\perp \cdot A'e^{i m_l \phi}{\sigmav_\pm})dz'\\
 = \int_{-\infty}^{z} e^{-i k z'}  \bigg{[}(\pm \widehat{\partial}_\rho A') -(\frac{m_l - i \widehat{\partial}_\phi}{\rho}  A')\bigg{]} e^{i (m_l \pm 1) \phi}dz' \label{DeltaE},
 \end{eqnarray} 
where $A' = A(\rho, \theta, z')$. Equation \ref{DeltaE} indicates that an electric field distinctly gains or loses unit topological charge $m_s = \pm 1$ when represented in the circular polarization basis.  The vortex phase coefficient $\exp[\pm i \phi]$ is interpreted as a path-dependent geometric phase \cite{Berry, Bliokh}, as it describes the phase that is accumulated around an azimuthal path in cylindrical coordinates, or a {\it star} singularity at a C-point, where the lines depicting pure circular polarization curve due to the finite transverse spatial extent of the electric field \cite{Berry2}.   Spin-orbit interactions arise because the spin and orbital angular momentum contributions, generally associated with the first and second round-bracketed terms of Eq. \ref{DeltaE}, respectively, add constructively to the longitudinal electric field {\it magnitude}.  Since the orbital angular momentum contribution scales inversely with radius, spin-orbit interactions are significant when the distances between phase singularities and scattering edges are less than the incident field wavelength.  
\begin{figure}[b]
\centering\includegraphics[width=3.5in]{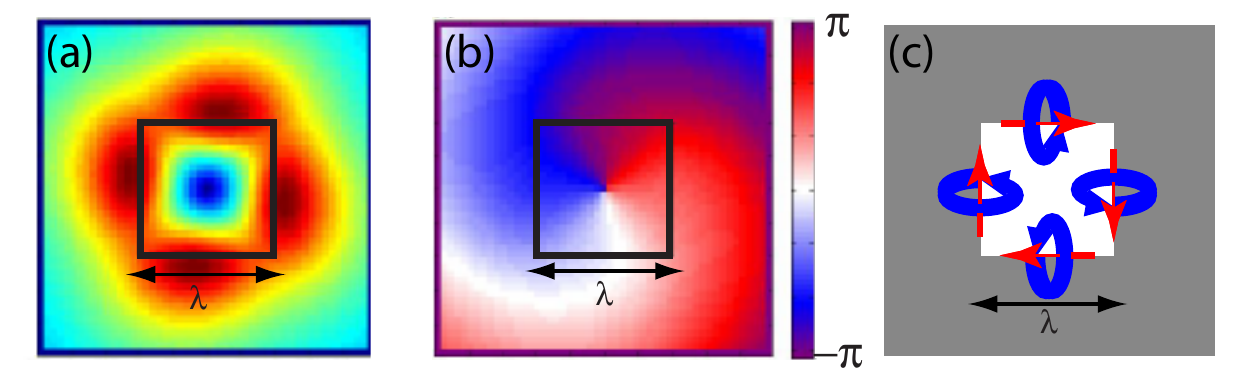}\caption{Numerically calculated (a) amplitude and (b) phase of the scattered longitudinal field produced by a left-handed or $\sigma_+$ circularly-polarized plane wave incident on a square metal aperture of width $L_x = L_y = \lambda$ in a metal sheet with thickness $\lambda/2$.  (c) Poynting vector (blue) and angular momentum vector (red) of the scattered longitudinal field.  The arrows reverse direction depending on the incident orthogonal-circular polarization illuminating the metal sample with square aperture.}\label{Fig1}
\end{figure}

We use an analytic mode solver \cite{Janne} to numerically calculate the fields transmitted through apertures in ideal metal sheets with cartesian symmetry.  Figure \ref{Fig1} illustrates the shadow-side longitudinal field at a distance $\lambda/10$ after the metal sheet of thickness $D = \lambda/2$, produced by a left-handed or $\sigma_+$ circularly-polarized plane wave with zero orbital angular momentum $m_l=0$ incident on a square aperture of length $L_x =L_y = \lambda$.  The amplitude [Fig. \ref{Fig1}(a)] demonstrates electric field enhancement at the aperture surfaces.  While the incident longitudinal field component has zero phase helicity $m_z=0$, the shadow-side phase contains a topological charge of $m_z = +1$ (clockwise, red-white-blue) [Fig. \ref{Fig1}(b)].  The orthogonal $\sigma_-$ or right-handed circularly-polarized field produces a vortex phase front with opposite topological charge $m_z = -1$ (clockwise, blue-white-red) [not shown].  Sharp aperture edges are responsible for the strong longitudinal field enhancement and large transverse field gradient.  The polarization-dependent vortex phase represents the complex exponential shown in Eq. \ref{DeltaE}.

If we decompose the total angular momentum into contributions in the longitudinal and azimuthal directions $<{\bf J}> = \rv \times ({\bf E} \times {\bf B^*}) +c.c.= J_z{\kv} + J_\phi \phiv$, then angular momentum in the  $z$-direction $J_z$ is considered a paraxial term and the second term $J_\phi$ involves spin-orbit interactions described here due to the longitudinal field components \cite{Barnett}.  The azimuthal component $J_\phi$ describes nonparaxial effects and changes sign with the incident circular polarization handedness.  In Fig. \ref{Fig1}(c) we illustrate the Poynting vector (blue arrows) and angular momentum vector (red arrows) of the scattered longitudinal field. 
\begin{figure}[t]
\centering\includegraphics[width=3.5in]{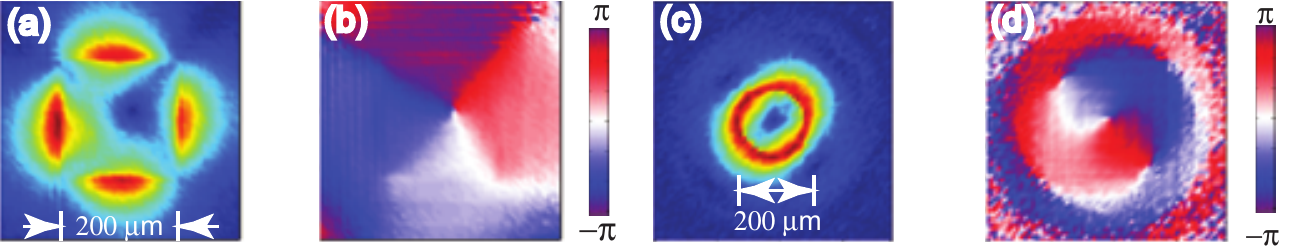}\caption{
The reconstructed amplitude and phase of scattered longitudinal fields produced by (a-b) square and (c-d) round apertures in aluminum film when illuminated with circularly-polarized light.  The incident light wavelength is 600GHz ($\lambda = 500\mu m$) and the aperture width and radius are approximately $200 \mu m$. The aluminum has a thickness of $80 \mu m$.}\label{Fig2}
\end{figure}

We experimentally measure the scattered longitudinal electric fields from subwavelength apertures using a THz near-field electro-optic detection method and focused probe beam, which provides full vector characterization of the transmitted shadow-side electric field with $10 \mu m$ resolution \cite{Planken}.  Scattering by circularly-polarized fields is accurately reconstructed using linearly-polarized THz signals.  Figure \ref{Fig2} shows transmitted longitudinal field amplitudes and phases produced by incident circularly-polarized light at 600 GHz ($\lambda = 500 \mu m$) on circular (radius $a=100 \mu m$) and square (length $L_x = L_y = 200 \mu m$) apertures.   Both square [Figs. \ref{Fig2}(a,b)] and round [Figs. \ref{Fig2}(c,d)] amplitudes show field enhancement due to interaction with aperture surfaces at the metal aperture edges, and the formation of an on-axis phase singularity.   Pairs of opposite-sign singularities appear off-axis in Fig. \ref{Fig2}(d), which we attribute to the non-normal angle of incidence between the signal and the sample.  Moreover, these off-axis phase singularities associated with misalignment change in location depending on the incident $\sigma_+$ or $\sigma_-$ orthogonal circular polarization.  

In the longitudinal component of electromagnetic fields, orbital angular momentum phase singularities {\it combine} with the polarization-dependent phase singularities associated with spin angular momentum and in the scattering by subwavelength apertures, the effects of these phase singularities is significant.  We consider input Laguerre-Gaussian profiles with index $p=0$
\begin{equation}
LG^{(m_l)}(\rho_m, \phi) = C\rho^{|m_l|} e^{-\rho_m^2/2} e^{i m_l \phi}\label{LG},
\end{equation}
where $C$ is a normalization constant such that $\int |LG^{(m_l)}|^2 dA = 1$, or $C = (\pi m_l!)^{-1/2}$, and the radial coordinate $\rho_m$ is normalized such that the mode field radius $(\int |LG^{(m_l)}|^2 \rho_m^2 dA)^{1/2} = 1$.  
\begin{figure}[t]
\centering\includegraphics[width=3.5in]{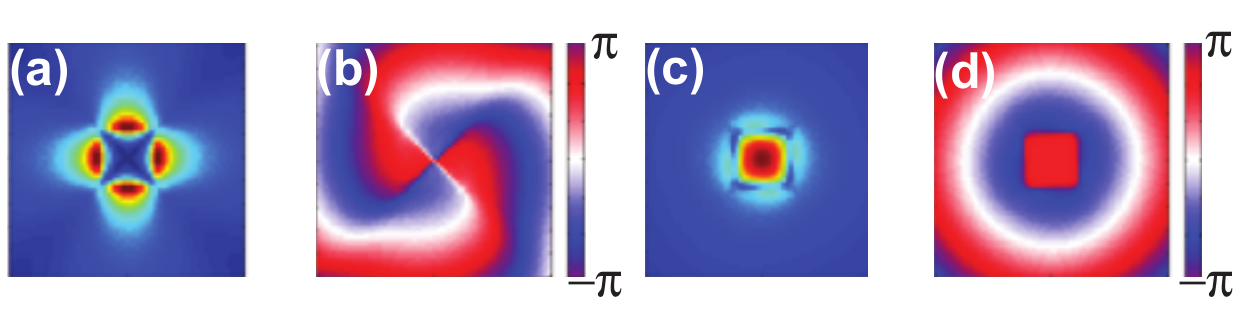}\caption{
Amplitudes and phases of the longitudinal $z$-component of transmitted fields produced by Laguerre-Gaussian beams of topological charge $m_0 = +1$ and (a-b) left-handed circular polarization and (c-d) right-handed circular polarization.}\label{Fig3}
\end{figure}

Figure \ref{Fig3} shows the amplitude and phase of the scattered-field $z$-components after transmission through a square aperture given incident orthogonal circularly-polarized Laguerre-Gaussian beams with topological charge of $m_l = + 1$.  The combination of left-handed circular polarization and the incident vortex phase produces a net transmitted topological charge of $m_z = +2$ [Fig. \ref{Fig3}(b)].  In contrast, the combination of right-handed circular polarization with similar phase vorticity cancel and produce a flat-phase $m_z = 0$ field [Fig. \ref{Fig3}(d)].  A comparison of the amplitudes indicate that right-handed circular polarization [Fig. \ref{Fig3}(c)] produces a shadow-side on-axis constructive maximum, whereas the amplitude of the left-handed circularly-polarized scattered field is strictly zero on-axis due to the on-axis phase singularities [Fig. \ref{Fig3}(a)].  Therefore, addition of a helical phasefront breaks chiral symmetry and the transmitted spatial beam profiles associated with orthogonal circular polarizations are no longer mirror images.

The coupling between spin polarization and orbital angular momentum phase fronts is observed in the total energy that is transmitted through apertures.  We use the result of \cite{Bouwkamp} to evaluate the transmission immediately after a circular sub-wavelength aperture within an infinitely-thin film of metal
\begin{equation}
E_z (\rho, \phi)= \frac{4\rho}{\pi (a^2-\rho^2)^{1/2}} [\cos\xi_0 \cos\theta_0 \cos\phi +\sin\xi_0 \sin\phi]\label{Bk2}
\end{equation}
where $\xi_0$ is the angle that measures between the electric field vector and the $x-z$ plane, $a$ is the radius of the aperture, $\rho$ and $\phi$ are the cylindrical coordinates of the scattered field, and the incident angle $\theta_0$ measures between the $z$-axis and the direction of incidence, ${\bf k}$.  Equation \ref{Bk2} provides an approximation of the scattered longitudinal fields for the condition $2\pi a/\lambda <1$. 
\begin{figure}[b]
\centering\includegraphics[width=3.5in]{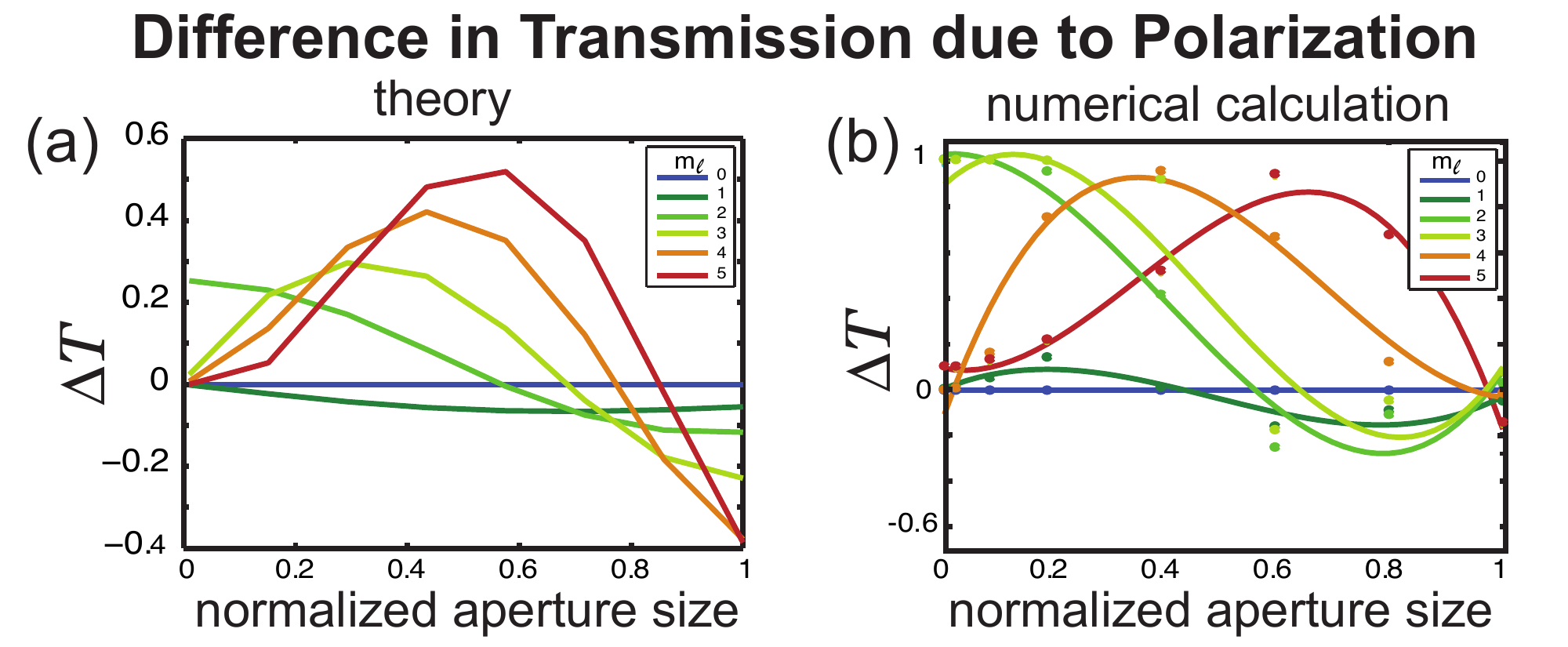}\caption{Difference in transmission between orthogonal-circularly polarized vortices $\Delta T = (T_{+}  -  T_-)/(T_{+} + T_{-})$ as a function of normalized hole aperture size for different incident topological charge $m_l$. (a) Theoretical prediction model given round aperture in infinitely-thing metal sheet. (b) Numerical calculation for a square aperture given finite sheet thickness $D=\lambda$.}\label{Fig4}
\end{figure}

We decompose incident Laguerre Gaussian beams [Eq. \ref{LG}] into plane waves with different wavevector ${\bf k}$ to solve for the scattering given by Eq. \ref{Bk2} and $\sigma_\pm$ circular-polarization
\begin{eqnarray}
E_{z}^{(m_l, \sigma_\pm)} (\rho_m, \phi)=\frac{C\rho_m e^{i(m_l\pm 1) \phi}}{\sqrt{a^2-\rho_m^2/(|m_l|+1)}}\times \nonumber \\
\int_{k_\rho = 0}^{k_\rho = k_c} (k_\rho)^{m_l} e^{-(k_\rho)^2/2} \times \nonumber \\
\bigg{[}(\kappa-1) J_{m_l\pm 2}(\rho_m k_\rho)- (1+\kappa)J_{m_l}(\rho_m k_\rho)\bigg{]} k_\rho\,\,dk_\rho \label{Eszk},
\end{eqnarray}
where the coefficient $\kappa=\sqrt{ 1-(k_\rho/k_m)^2}$ contains the normalized wavenumber $k_m = 2\pi / \lambda \sqrt{|m_l|+1}$, and the cutoff transverse wavenumber is $k_c$. For this investigation, we consider that plasmons are excited on the incident metal surface so that $k_c =\infty$; a cutoff wavenumber of $k_c=k_m$ implies that, for example, a dielectric coating prevents surface waves or plasmons from propagating on the incident metal surface.  This ${\bf k}$-space relation, Eq. \ref{Eszk}, indicates that the scattered field amplitudes couple into Bessel functions of order $m_l$ and $m_l\pm 2$, while the longitudinal phase vorticity remains described by the relation $m_z = m_l + m_s = m_l \pm 1$.  The electromagnetic spin-orbit interaction exists in the Bessel term $J_{m_l \pm 2}$, which indicates that the energy transmitted through the aperture depends on both spin angular momentum and orbital angular momentum.

In Fig. \ref{Fig4} we show the difference in transmission associated with each orthogonal polarization $\Delta T = (T_{+}  -  T_-)/(T_{+} + T_{-})$ for varying orbital angular momentum topological charge $m_l$ as a function of aperture diameter or width $L_x=L_y=2a$, where the transmission is $T_\pm = \int |E_z^{(m_l, \sigma_\pm)}|^2 dA$.  The aperture width is normalized by the input beam mode field diameter and since there is no cutoff transverse wavenumber $k_c = \infty$, our analytical calculation is independent of wavelength.   Theoretical predictions are shown in Fig. \ref{Fig4}(a).  In Fig. \ref{Fig4}(b), we plot the difference in transmission from numerical simulations for metal sheet thickness $D = \lambda$ and polynomial curve-fit lines.  The difference in transmission $\Delta T$ represents the spin-orbit interaction via electromagnetic scattering and is therefore zero when there is no orbital angular momentum or $m_l = 0$.  

The difference in transmission $\Delta T$ for a single incident vortex $m_l=1$ is less than 10\%, while that for higher-order vortices exceeds 50\% depending on aperture size.  Both numerical calculations and theoretical analysis predict similar aperture sizes for maximum spin-orbit interaction or maximum $|\Delta T|$.  We observe a maximum spin-orbit interaction for $m_l=2$ when aperture sizes approach zero, and the maximum spin-orbit interactions occur for increasing $m_l$ at increasing aperture widths.  Our theoretical prediction strongly underestimates the difference in transmission, particularly at small aperture sizes, and this is not reconciled by changing the metal sheet thickness in numerical simulations.   Discrepancies arise from the comparison between cylindrical (theoretical) and cartesian (numerical) symmetry, however this point does not entirely resolve the differences described.  

Although the maximum change in the longitudinal field topological charge $m_z$ due to polarization and scattering at a single aperture is $m_s = \pm 1$, we demonstrate our ability to manipulate and control spin-orbit interactions by producing shadow-side fields with arbitrary phase vorticity.  The initial conditions that produce fields of arbitrary phase fronts are not uniquely defined unless there exists constraints placed on the amplitude of the scattered fields.  We use a ``necklace'' arrangement of $n$ apertures, where $n$ corresponds to the desired phase vorticity $m_z$, and prepare the input field polarization as a coherent superposition of azimuthally $ {\bf \widehat {e}_{AP}}$ and radially $ {\bf \widehat {e}_{RP}}$ polarized fields, or
\begin{eqnarray}
\widehat{\bf e}_\pm^{RA}(\phi) =\frac{1}{\sqrt{2}}[ {\bf \widehat{e}_{RP}} (\phi)\pm i  {\bf \widehat {e}_{AP}}(\phi)].\label{Easp}
\end{eqnarray}
There is zero angular momentum in a $\widehat{\bf e}_\pm^{RA}$-polarized field that has a radially-symmetric field amplitude, although $\widehat{\bf e}_\pm^{RA}$ carries phase singularities that are visible when decomposed into the circular-polarization basis, or ${\bf \widehat {e}_{\stackrel{RP}{AP}}}(\phi)  = [\exp({-i\phi})\sigmav_+ \pm \exp({i\phi}) \sigmav_-]/\sqrt{2}$.
 
\begin{figure}[t]
\centering\includegraphics[width=2.5in]{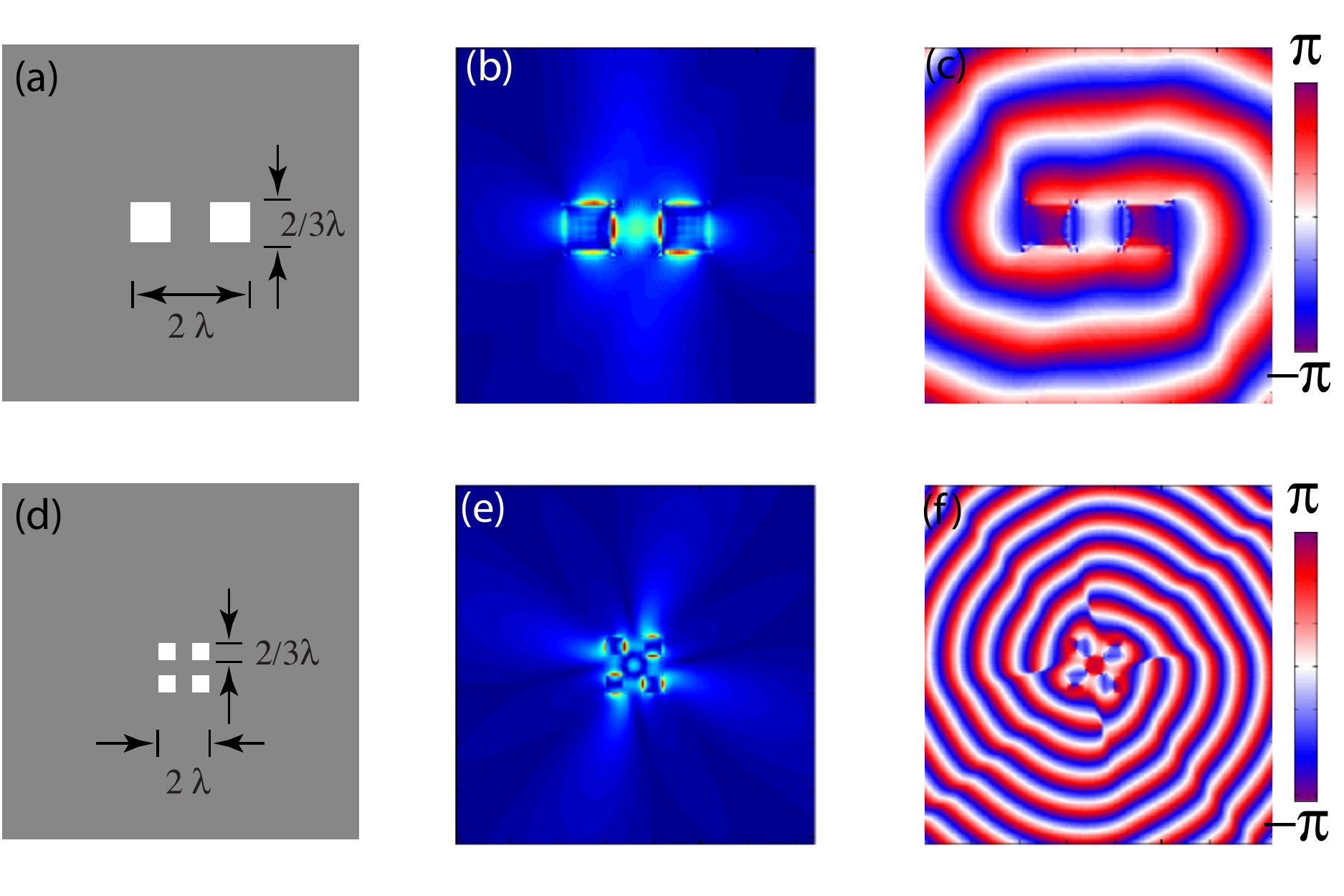}\caption{(a) Two apertures with dimensions and separation $\Delta x = \Delta y = L_x=Ly=2/3 \lambda$ and correspondingly scattered (b) amplitude and (c) phase produced via numerical calculations.  (d)Four apertures with dimensions and separation $\Delta x = \Delta y = L_x=Ly=2/3 \lambda$ and the corresponding scattered (e) amplitude and (f) phase produced via numerical calculations.  The metal sheet has thickness $D = \lambda/4$. }\label{Fig5}
\end{figure}

With a $\widehat{\bf e}_\pm^{RA}$-polarized field centered and incident on an azimuthal arrangement of $n$ apertures, each aperture produces a single topological charge whose sign is determined by the local $\pm$-handedness in Eq. \ref{Easp}, and the total charge associated with the entire scattered field is $m_z=n m_s$.  In Figs. \ref{Fig5}(a) and \ref{Fig5}(d) we show $n=2$ and $n=4$ equally-spaced square holes separated by and with dimensions $\Delta x = \Delta y = L_x=Ly=2/3 \lambda$.  We numerically calculate the transmission assuming a metal sheet thickness $D=\lambda/4$.   Figures \ref{Fig5}(b,e) show the amplitudes and Figs. \ref{Fig5}(c,f) illustrate the phases that demonstrate singularities $m_z = + 2 m_s$ and $m_z = + 4 m_s$.  Incident fields have Laguerre-Gaussian magnitudes $|LG^{(m_l = 1)}|$, although we observe that the amplitude of the incident field spatial beam profile does not substantially affect the phase of the scattered fields.

In conclusion, electromagnetic spin-orbit interactions that occur via scattering are significant when vortex phase singularities associated with orbital angular momentum propagate at subwavelength distances to scattering edges.  Here we have demonstrated that phase singularities associated with circular polarization and vortex phase fronts combine in the longitudinal electric field components.  With the addition of orbital angular momentum, the energy that is scattered through round and square subwavelength apertures is polarization-dependent.  Depending on the aperture size and incident phase vorticity, the difference in transmission due to polarization can exceed 50\%.  Our research indicates that planar asymmetric or chiral metal nanostructures that change the polarization of scattered fields also impart a phase vorticity and this explains previously observed polarization-dependent spatial beam profiles \cite{ChiralHole1, ArchSpiral, ChiralHole2}.  Our results illuminate new considerations for manipulating plasmons or surface waves, and designing and aligning metamaterials.  

The author graciously acknowledges funding from the Fulbright and Netherlandic-American Foundations via Philips Electronics North America. 


\thebibliography{}
\bibitem{NFMeasure} M.L.M. Balistreri {\it et al.}, Phys. Rev. Lett. {\bf 85}, 294 (2000).
\bibitem{NFSing} A. Nesci {\it et al.}, Opt. Comm.  {\bf 205}, 229 (2002).
\bibitem{PIO} M.S. Soskin and M.V. Vasnetsov (2001) Amsterdam Elsevier, Progress in Optics, {\bf 42}, 219 (2001); L.J. Allen {\it et al.}, Amsterdam Elsevier, Progress in Optics, {\bf 42}, 219 (2001).
\bibitem{Berry2} M.V. Berry and M.R. Dennis,  Proc. R. Soc. Lond. A {\bf 457}, 141 (2001); M. R. Dennis, Opt. Comm {\bf 213}, 201 (2002).
\bibitem{Berry3} M.V. Berry and M.R. Dennis, J. Phys. A {\bf 40}, 65 (2007); R.W. Schoonover and T.D. Visser, Opt. Express {\bf 14}, 5733 (2006).
\bibitem{microparticle} H. He {\it et al.}, \prl {\bf 75}, 826 (1995).
\bibitem{ChiralHole1} A. Papakostas {\it et al.}, Phys. Rev. Lett. {\bf 90}, 107404 (2003); Kuwata-Gonokami {\it et al.}, Phys. Rev. Lett. {\bf 95}, 227401 (2005).
\bibitem{ArchSpiral} T. Ohno and S. Miyanishi, Opt. Express {\bf 14}, 6285 (2006).
\bibitem{ChiralHole2} A.V. Krasavin {\it et al.}, J. Opt. A {\bf 8}, S98 (2006).
\bibitem{WhirlPart} M.V. Bashevoy {\it et al.}, Opt. Express {\bf 13}, 8372 (2005).
\bibitem{LenstraPRL} H.F. Schouten {\it et al.}, Phys. Rev. Lett. {\bf 93}, 173901 (2004). 
\bibitem{Leuchs} J. Kindler {et al.,} App. Phys. B {\bf 89}, 517 (2007).
\bibitem{Janne} J.M. Brok and H.P. Urbach, Opt. Express {\bf 14}, 2552 (2006).
\bibitem{Zeldovich} V.S. Liberman and B.Ya. Zel'dovich, \pra {\bf 46}, 5199 (1992).
\bibitem{Chiao} R.Y. Chiao {\it et al.}, Phys. Rev. Lett. {\bf 60}, 1214 (1988).
\bibitem{HiNAPRL} Y. Zhao {\it et al.}, Phys. Rev. Lett. {\bf 99}, 073901 (2007); K. Lindfors {\it et al.}, Nat. Photonics {\bf 1}, 224 (2007).
\bibitem{Berry} M.V. Berry, Proc. R. Soc. Lond. A {\bf 392}, 45 (1984). 
\bibitem{Bliokh} K.Y. Bliokh, \prl {\bf 97}, 043901 (2006). 
\bibitem{Barnett} S.M. Barnett, J. Opt. B {\bf 4}, S7 (2002).
\bibitem{Planken} A.J.L. Adam {\it et al.}, Opt. Express {\bf 16}, 7407 (2008). 
\bibitem{Bouwkamp} C.J. Bouwkamp, Rep. Prog. Phys. {\bf 17}, 35 (1954).
\end{document}